\documentclass{appolb}
\usepackage{graphicx}
\usepackage{amsmath}

\begin{document}

\title{Efficiency of pair formation in a model society
\thanks{Presented at the 2nd Polish Seminar on Econo- and Sociophysics, Cracow, 21-22 April 2006}
}
\author{M. Wasko
\footnote{\tt wajs1@o2.pl}
\address{Faculty of Physics and Applied Computer Science,
AGH University of Science and Technology,
al. Mickiewicza 30, PL-30059 Cracow, Poland}
\and
K. Ku{\l}akowski
\footnote{\tt kulakowski@novell.ftj.agh.edu.pl}
\address{Faculty of Physics and Applied Computer Science,
AGH University of Science and Technology,
al. Mickiewicza 30, PL-30059 Cracow, Poland}
}
\maketitle

%% \date{\today}

\begin{abstract}
In a recent paper a set of differential equations was proposed to
describe a social process, where pairs of partners emerge in a community. 
The choice was performed on a basis of attractive resources and of 
random initial preferences. An efficiency of the process, defined as
the probability of finding a partner, was found to depend on the community 
size. Here we demonstrate, that if the resources are not relevant, the 
efficiency is equal to unity; everybody finds a partner. With this new 
formulation, about 80 percent of community members enter into dyads; 
the remaining 20 percent form triads.
\end{abstract}


%%\keywords{Agents ; Dynamics ; Marriage ; Resources}

\section{Introduction}

Sociophysics is a branch of the statistical and computational physics. It can be defined
as an attempt to apply tools of theoretical physics to the social sciences. 
The problem, if and to which extent it is possible/allowed to describe social processes
mathematically, cannot by resolved a priori. The arguments for the negative answer are 
well known in sociology, as they are used in the discussion between the empirical and the 
theoretical traditions \cite{ox}. As these arguments are continuously developed during
about 200 recent years, it seems possible that still the subject is not closed. As for 
the physicists, however, mathematics is at least not worse than any other language;
then the discussion is empty. Last but not least, the empirical tradition embraces a 
good part of sociology.  In this field, however, it is natural to treat the physicists as
outsiders, who "reinvent existing tools and rediscover established empirical results" 
\cite{free,russ}, and who eventually will be integrated into the larger social network 
community. We imagine that as for the sociological perspective, these notions 
could be applied to the sociophysics as a whole. Feeling offended or not, we are not willing 
to defy the integration.

The problem to be discussed here can be strictly formulated as a set of abstract differential 
equations, but its social interpretation is maybe more appealing. The term we use: "pair formation" 
means that a set of $N$ objects, initially random, transforms in time towards 
a stable state with some pairs of objects. If no object remains unpaired, the efficiency of the process
is 100 percent or 1.0. A fraction of unity means that some objects are not paired,
i.e. the process is less efficient. The most straightforward equivalent of the process
is when pairs of sexual partners appear in a group of young people. However, as we are
going to demonstrate, the mathematics involved does not contain an assumption on 
individual differences, as for example sex.

Up to our knowledge, the problem of social pairing was discussed in Refs. \cite{cald,smit}. The
subject of these papers is some kind of social optimization: the task is to find a partner with
best parameters. There, the evaluation of these parameters can take into account individual needs.
This algorithmic approach is well established in the literature of the subject \cite{nash,gusf}.
Our starting point is maybe different, as we base on an assumption that final equilibrium state 
is developed as a result of initial preferences which are random. In this picture, the preferences
are first followed, then rationalized ex-post. We note that this point of view was explored 
recently in a model of random sexual contacts \cite{glh}. As it was formulated by Vilfredo Pareto, 
people do what they believe is desirable for them \cite{par}. 

In our previous paper \cite{karp}, an attempt was made to evaluate the efficiency of the 
pairing process as dependent on the size $2N$ of the whole group (this variable was termed $N$ 
in \cite{karp}). The difference between Ref. \cite{karp} and the present work is in the details
of the model equations; in words, it can be summarized as follows. The core of the pairing process
is that each group member selects another one based on the recources of the selected member
and on her/his willingness to share them. Further, the willingness of the selecting member to share
her/his own resources is reoriented in time as to give most resources to the most generous member,
just selected. This can be seen as an application of the concept of reciprocity, well established 
in social sciences \cite{mma,axe,sch}. In this way, a positive feedback appears between particular 
members and weak initial
inclinations are transformed into a monopoly on mutual feeding. Those who do not reorient 
quickly enough, because their initial inclinations were ill-directed or too distributed - fail
and remain unpaired. This version of the model has been presented in Ref. \cite{karp}. The 
velocity $\alpha $ of reorientation was another parameter, and the efficiency was investigated
as dependent on $\alpha $ and $N$.

Within the above given terms, the goal of this work can be presented as follows. We already know 
that an individual selection is based on two agents: resources of the selected partner and her/his 
willingness to share them with the selecting member. Now, let us eliminate the criterion
of resources, with only the one of willingness left. In other words, it does not matter any
more if a member has anything to offer; the only relevant question is to whom she/he wants to offer what 
she/he has. Then, a striking effect is observed: the efficiency of the pairing process becomes equal
to unity. No unpaired members is left.

The model equations are given in the next section, together with some example of the obtained
plots. Last section contains our conclusion, aimed as sociological - in the physical sense.

\section{Model and its results}

The resources of $i$-th unit is defined as $p(i)\ge 0$. It evolves
according to the following rule.

\begin{equation}
\frac{dp(i)}{dt}=N^2-[\sum_{i=1}^Np(i)]^2-\sum_{j\ne
i}^N[r(j,i)p(j)-r(i,j)p(i)] \end{equation}
where $t$ is time and $r(i,j)$ is what $i$-th unit indends to offer to $j$-th one. The matrix $r(i,j)$ 
also evolves; in the model version presented in Ref. \cite{karp} it was evolving
according to the following rule:

\begin{equation}
\frac{dr(i,j)}{dt}=\alpha \Big(r(j,i)p(j)-\frac{\sum_k r(k,i)p(k)}{N-1}\Big)
\end{equation}

whereas in the present work the latter equation is limited to

\begin{equation}
\frac{dr(i,j)}{dt}=\alpha \Big(r(j,i)-\frac{\sum_k r(k,i)}{N-1}\Big)
\end{equation}

As we see, in the previous model the matrix  $r(i,j)$ (willingness matrix) evolves in time as dependent
on the resources $p(i)$. In the second formulation this dependence is removed.

In Fig. 1 we present an example of the results for small value of $\alpha $. In the former
version of the model \cite{karp}, the efficiency of the process for this value of $\alpha $ 
was about 0.2. With the application of Eq.3, the efficiency is equal to 1.0 (100 percent) for
all of our numerical evidence. As in the previous model, the paired units get the same amounts
of resources.

\begin{figure}
\begin{center}
\resizebox{0.80\hsize}{!}{\includegraphics{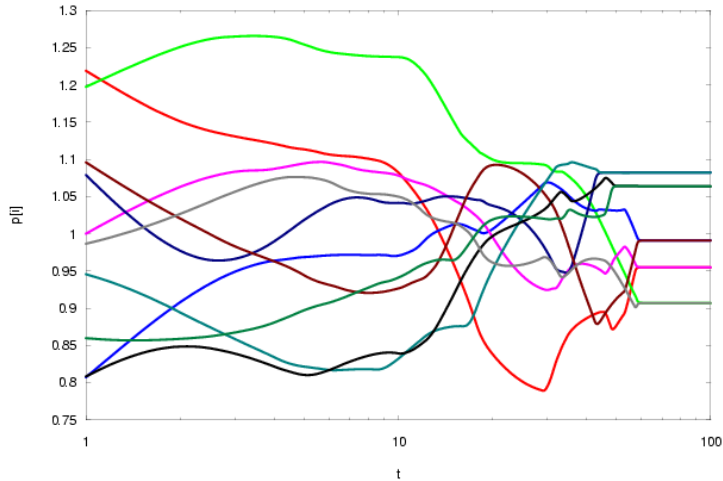}}
\end{center}
\caption{\label{fig-plot} The amount of resources as dependent on time for $N$=10 and $\alpha$=0.2.}
\end{figure}

Surprisingly, a pair is not the only pattern which appears as the result of the simulation. We observe
also triads of two different types; let us call them type 1 and 2. In the stable state of triad 1,
one of contributing units exchanges resources symmetrically with the remaining two, while these two
do not interact with each other. As a consequence, the amount of resources of one unit is twice larger than the 
amounts of the others. In triad 2, each of three units exchanges her/his resources with the others; then, their
amounts of resources are equal. These triads are presented schematically in Fig. 2, together with the pair.
We checked that the probability that an unit enters to a triad is about 0.2, and this result seemingly does
not depend on the parameters $N$ and $\alpha$. Small numerical evidence suggests that groups larger than 
triads can be stable as well, but they are rare.

\begin{figure}
\begin{center}
\resizebox{0.50\hsize}{!}{\includegraphics{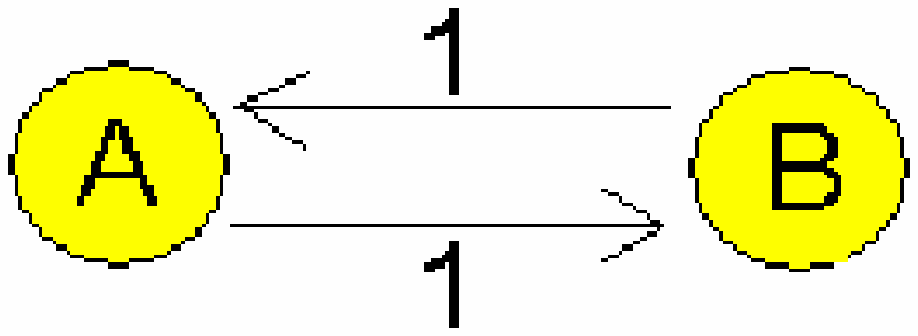}}
\resizebox{0.50\hsize}{!}{\includegraphics{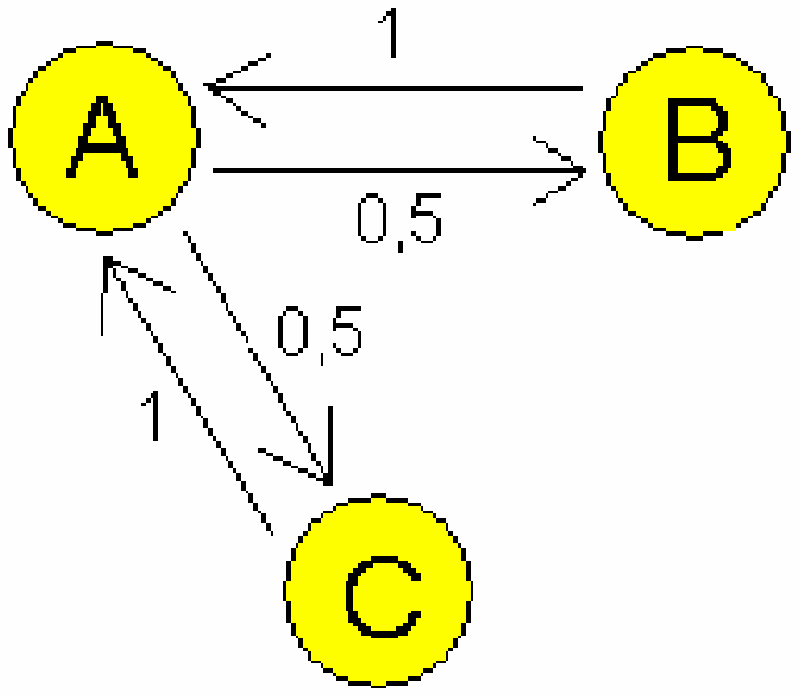}}
\resizebox{0.50\hsize}{!}{\includegraphics{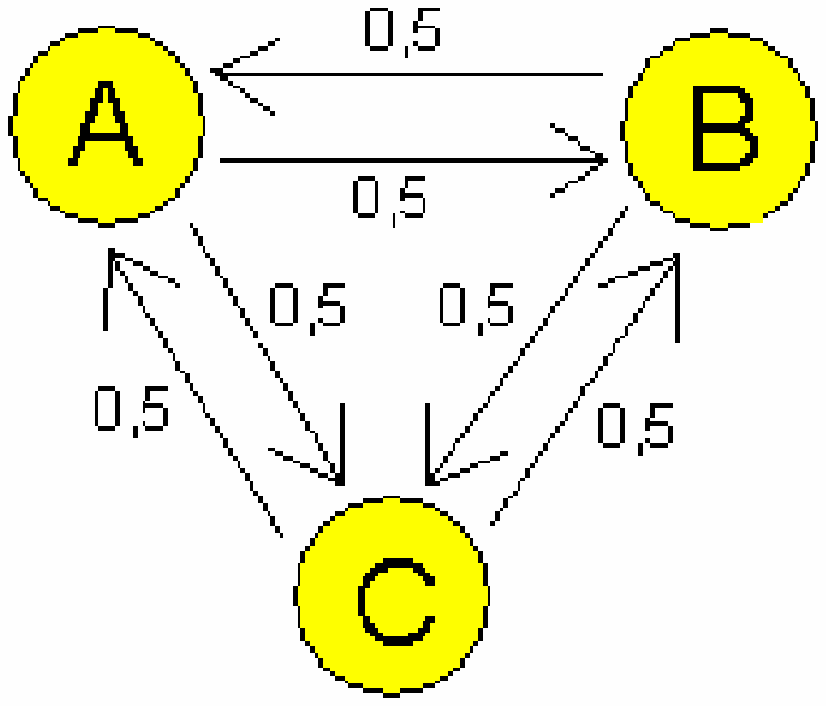}}
\end{center}
\caption{\label{fig-pairs} The scheme of a pair and two kinds of triads.}
\end{figure}

\section{Conclusion}

Summarizing, we report a new strategy of searching a partner to exchange resources. Our new element is to
disregard the actual amount of resources of a potential partner, and to take into account only
her/his willingness to offer them. Although this strategy is obviously far from being optimal with respect 
of acquired amount of resources, its advantage is that everybody is fed with some resources in the final 
state. This is the main goal of this paper.

The results of the model seem to present a nontrivial alternative for what is considered as normal
behaviour in biology. Sexual rivalization and competition of genes - these ideas entered into 
common knowledge. The price we pay - we, living creatures - is that some less perfect units cannot 
pass their genetic material to an offspring. In the competitive world and in a closed group, they just are
not able to find partners, tolerant to their weaknesses. Although in human world all that can be 
changed at least in principle, the literature and other media provide a large pile of descriptions
of this situation.

Would it then be of interest to modify our sexual behaviour? Truly, we do it already. As it is described 
by evolutionary psychologists, once the aim is not an offspring but just sex, our demands decrease
\cite{buss}. This behaviour means that our sexual needs are alienated from their biological basis.
Then, they can be used for other purposes: political, economic, social or just entertainment. This
is one of differences between animals and people.

\section*{acknowledgments} Thanks are due to our Anonymous Referee for helpful remarks, which allowed 
to introduce the references data on the reciprocity principle.

\end{document}